\documentclass[sigconf]{acmart}

\setcopyright{none}  
\renewcommand\footnotetextcopyrightpermission[1]{}

\acmConference[Augmented Educators and AI(CHI 2025 Workshop)]{Augmented Educators and AI: Shaping the Future of Human-AI Collaboration in Learning on CHI 2025 Workshop}{April 26,2025}{Yokohama, JAPAN}

\begin{document}

\title{AI for Accessible Education: Personalized Audio-Based Learning for Blind Students}

\author{Crystal Yang}
\email{crustalyang@gmail.com}
\orcid{1234-5678-9012}
\affiliation{%
  \institution{University of Pennsylvania (incoming), Obra D. Tompkins High School}
  \city{Katy}
  \state{Texas}
  \country{USA}
}

\author{Paul Taele}
\email{ptaele@tamu.edu}
\affiliation{%
  \institution{Texas A\&M University}
  \city{College Station}
  \country{USA}
}

\renewcommand{\shortauthors}{Yang et al.}

\begin{abstract}
Blind and visually impaired (BVI) students face significant challenges in traditional educational settings. While screen readers and braille materials offer some accessibility, they often lack interactivity and real-time adaptability to individual learning needs. This paper presents Audemy, an AI-powered audio-based learning platform designed to provide personalized, accessible, and engaging educational experiences for BVI students. Audemy uses adaptive learning techniques to customize content based on student accuracy, pacing preferences, and engagement patterns. The platform has been iteratively developed with input from over 20 educators specializing in accessibility and currently serves over 2,000 BVI students. Educator insights show key considerations for accessible AI, including the importance of engagement, intuitive design, compatibility with existing assistive technologies, and the role of positive reinforcement in maintaining student motivation. Beyond accessibility, this paper explores the ethical implications of AI in education, emphasizing data privacy, security, and transparency. Audemy demonstrates how AI can empower BVI students with personalized and equitable learning opportunities, advancing the broader goal of inclusive education.
\end{abstract}


\maketitle

{\small
\noindent ©  This paper was adapted for the \textit{CHI 2025 Workshop on Augmented Educators and AI: Shaping the Future of Human and AI Cooperation in Learning},
held in Yokohama, Japan on April 26, 2025. This work is licensed under the Creative Commons Attribution 4.0 International License (CC BY 4.0).
}

\hyphenpenalty = 9000
\section{Introduction}

Blind and visually impaired (BVI) students face challenges in traditional classrooms due to the reliance on visual instructional materials and limited adaptive learning tools \cite{huff2023perceptions}. While screen readers and braille materials provide some accessibility, they often fail to offer interactive, real-time, and emotionally responsive learning experiences \cite{klingenberg2020digital}. Studies indicate that BVI students benefit most from multi-sensory learning environments, integrating auditory, tactile, and interactive digital technologies \cite{steinbach2022looking}, but existing accessibility tools lack the ability to adapt based on individual student progress and engagement \cite{essa2023personalized}. 

Given these challenges faced by BVI students, it is crucial to develop adaptive and accessible AI-driven educational tools that integrate into teaching practices. Therefore, we propose \href{https://audemy.org}{Audemy}, an AI-powered audio-based learning platform designed for blind and visually impaired (BVI) students, incorporating conversational AI and adaptive learning. In just one year, Audemy has grown to serve over 2,000 BVI students and has been iteratively developed through collaboration with more than 20 educators that specialize in accessibility and personalized AI learning. 

This paper explores insights about how augmented AI educators are able to deliver personalized, adaptive, and empathetic instruction by:
\begin{enumerate}
    \item Aligning games features with feedback from educators for BVI students.
    \item Adjusting AI-human interactions based on real-time responses of BVI students.
    \item Addressing ethical concerns in AI-powered education, ensuring privacy, equity, and human oversight in adaptive learning environments.
\end{enumerate}

Through analyzing user feedback from students and teachers gained in the development and commercialization process of Audemy, this paper contributes to best practices for creating accessible AI for inclusive education to support BVI students and their teachers.

\section{Literature Review}
Previous works have existed that attempt to create AI-driven education technologies for BVI students. 

AI-powered assistive technologies have significantly improved access to education for BVI people through facilitating interaction with digital content. For example, screen readers such as JAWS and NVDA use AI-powered text-to-speech synthesis to enable BVI students to navigate digital environments ~\cite{das2022design}. Similarly, Intel has supported advancements in AI for accessibility through initiatives like the AI for Youth program and collaborations that empower developers to create inclusive solutions. This contributes to tools that help blind and visually impaired students in real-time learning environments.

Adaptive learning platforms can dynamically adjust content difficulty, pacing, and modality ~\cite{yaseen2025impact}. Dyslexia-friendly fonts and Proloquo2Go improve literacy for students with visual impairments and learning disabilities by optimizing text readability and communication methods ~\cite{bachmann2018dyslexia}. Notability integrates speech-to-text AI for accessible note-taking, benefiting both BVI students and students with dyslexia. These tools demonstrate how AI facilitates customized learning environments that address diverse educational needs.

Conversational AI has shown promise in improving learning engagement for BVI students through interactive dialogue-based learning, and has been used to provide real-time feedback and adaptive questioning ~\cite{huq2024dialogue,yang2023heard}. In addition,  the effectiveness of AI-driven dialogue systems in accessibility-focused education ~\cite{mina2023leveraging}.

\section{Adaptive Audio-Based Learning for Blind Students}

\subsection{Feature Design Process}
Many design decisions in Audemy were developed via teacher and student feedback with the priority of helping students remain motivated and engaged. The features in Audemy were implemented as a result of direct feedback from teachers working with BVI students. 20 educators who specialize in teaching BVI students were interviewed to identify specific educational needs. The school affiliations of the teachers can be found in Table 1. 

\begin{table}[h]
    \centering
    \begin{tabular}{l c}
        \toprule
        \textbf{School Name} & \textbf{Count} \\
        \midrule
        Texas School for the Blind and Visually Impaired & 6 \\
        Kansas State School for the Blind & 2 \\
        Maryland School for the Blind & 1 \\
        California School for the Blind & 3 \\
        Illinois School for the Visually Impaired & 1 \\
        HeZe Special Education Center & 3 \\
        Arizona State School for the Deaf and the Blind & 4 \\
        \bottomrule
    \end{tabular}
    \caption{Counts of Students at Various Schools for the Blind and Visually Impaired}
    \label{tab:schools}
\end{table}

The iterative process consists of cycles of research, development, testing, and improvement to make sure that the platform remains effective and aligned with student needs, as shown in figure 1. This development workflow was supported by Intel, whose AI PC tools and hardware enabled efficient prototyping and testing of AI-driven features. All quotes in the paper are from interviewed teachers.

\begin{figure}[H]
    \centering
    \includegraphics[width=0.4\textwidth]{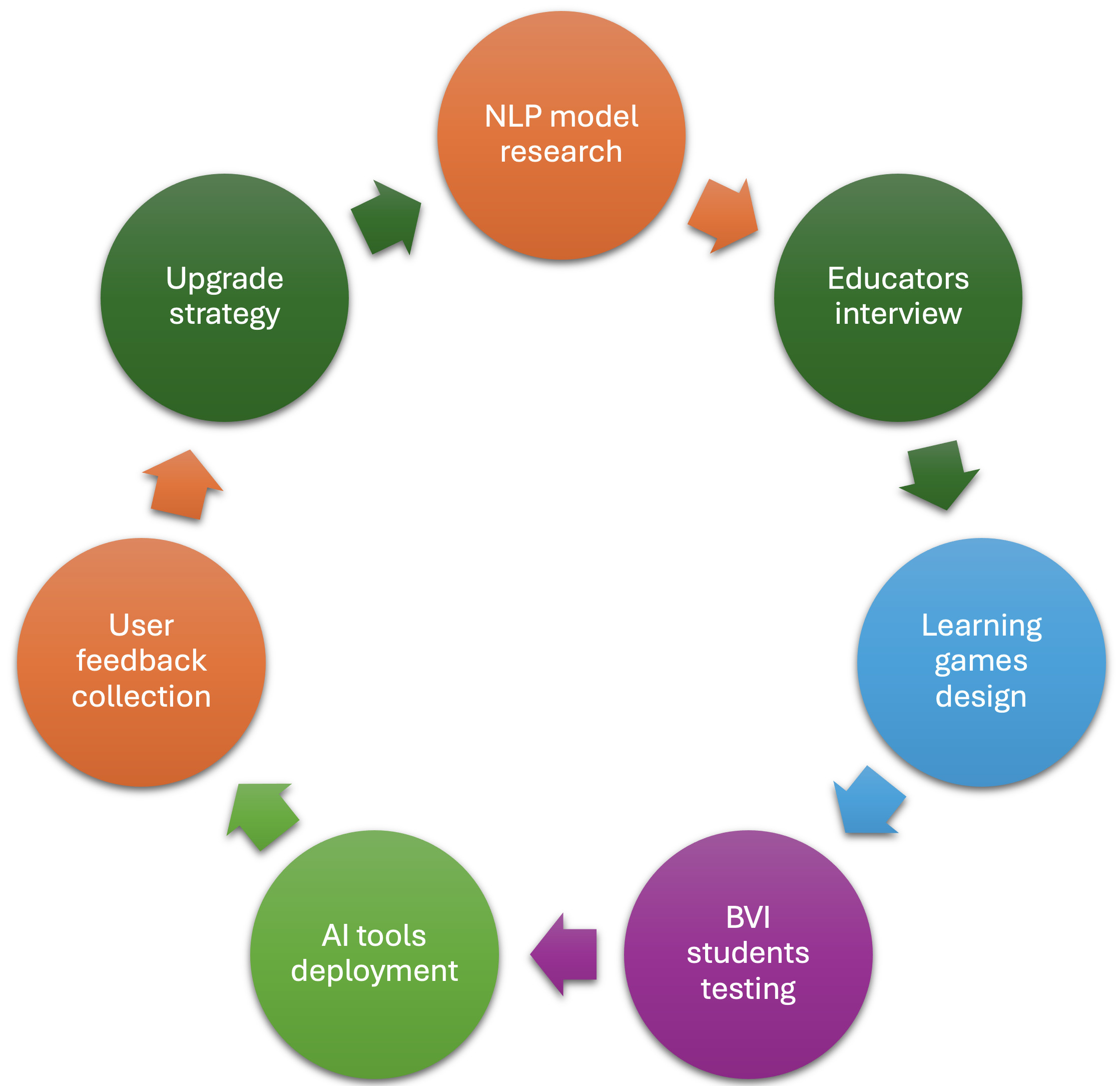}
    \caption{Development Feedback Loop}
    \Description{A diagram showing the development feedback loop.}
\end{figure}

\subsection{Platform Interactions Hierarchy}
Audemy’s interaction hierarchy ensures an accessible and intuitive learning experience for BVI students. The platform starts with an optimized, screen-reader-compatible interface, followed by an easy-to-navigate game selection process designed to minimize cognitive load while maintaining engagement. Audio-based customization options allow students to adjust playback speeds, enhancing usability and personalization.

To support adaptive learning, Audemy dynamically adjusts question difficulty based on student responses and provides real-time progress tracking. Educators benefit from AI-driven insights, helping them tailor instruction to individual student needs. Continuous updates introduce new educational games and system enhancements, keeping the learning experience dynamic and responsive. This structured approach ensures that accessibility, engagement, and adaptability remain central to Audemy’s design. Throughout this development, we utilized Intel-powered AI PCs to run all AI functionalities locally, AI processing is conducted locally on-device using Intel-powered systems, avoiding cloud transmission of sensitive voice or learning data. This setup also allowed for fast prototyping and privacy-conscious deployment without relying on cloud-based processing, which is particularly important when working with sensitive educational data, as it reduces privacy risks and aligns with principles of ethical AI in educational settings, particularly when working with minors.

\begin{figure}[H]
    \centering
    \includegraphics[width=0.4\textwidth]{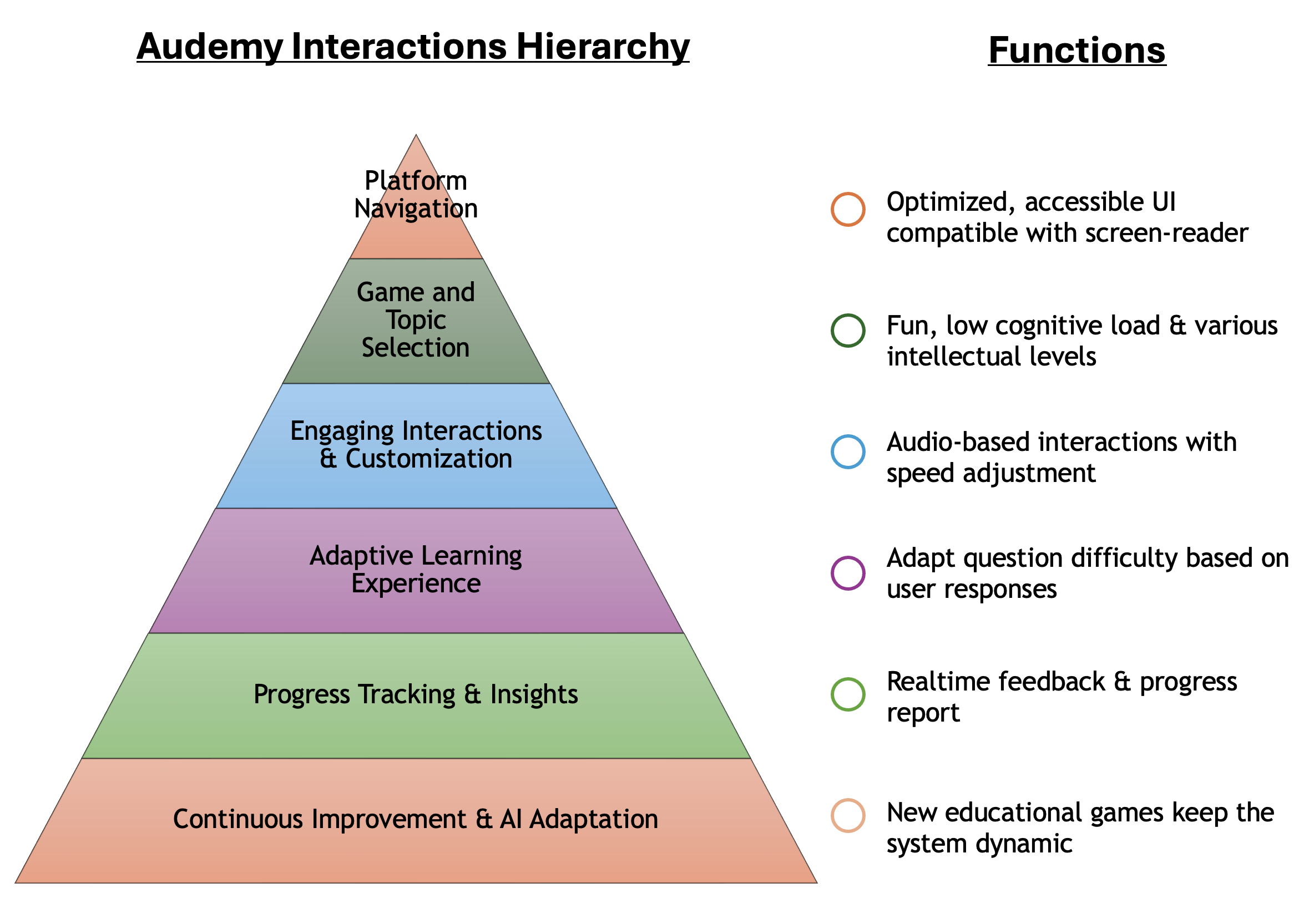}
    \caption{Audemy Interactions Hierarchy and key functions}
\end{figure}

\subsection{Adaptive Learning Features for Blind Students}
Through the feature design process, we have implemented 3 key adaptive learning features for blind students. Key adaptive learning features include:
\begin{itemize}
    \item \textbf{Adjusts question difficulty} based on student accuracy. This feature was built because many teachers talked about the importance of keeping students challenged while avoiding frustration, for example a teacher specializing in early-age blind education (T1) "if a student gets too many questions wrong, they disengage. But if it's too easy, they don't grow. The adjusting difficulty keeps them in that ideal learning zone.”
    \item \textbf{Modifies content delivery speed} according to repeat function usage. This feature was built because teachers noticed that some students needed extra time to process information, while others students preferred a faster pace. T1 stated "some students need repetition, but others pick things up quickly and get frustrated when forced to go at the same pace."
    \item \textbf{Presents content in diverse formats} (e.g., direct questions vs. real-world applications). This supports different learning styles. For example, a math game can present a basic equation like “What is 7+5?” or a contextual problem like “There are 7 bananas and 5 oranges. How many fruits are there in total?” This feature was built because some teachers talked about the need for multiple types of engagement. A Pre-K teacher for blind and visually impaired students (T2) stated "some students thrive with straightforward drills, while others need real-world context to grasp concepts."
\end{itemize}

\subsection{Practical Considerations for Accessible Learning AI}
To understand accessible and engaging design, we conducted a thematic analysis of the transcripts from teacher interviews. The outcomes led to a total of 25 codes, which were then grouped into 10 derived categories and 5 emergent themes. The following quotes illustrate and support the key themes and considerations that emerged as prevalent throughout our analysis:

\begin{itemize}
    \item Tools should \textbf{maximize engagement}. T1 stated “a crucial concept is to make learning fun. Basically, we need tools that grab their attention.” AI should use storytelling, gamification, and dynamic interaction to keep up learning retention.
    \item Interfaces should be kept \textbf{simple}. An accessibility specialist (T3) stated that “good sound and clear instructions” are important for students to use the platform independently. 
    \item Systems must \textbf{work with assistive technologies} that BVI students already use. T3 stated “[tools] have to work with screen readers and other tools they already use.” For example, screen readers and zoom text tool compatibility is needed. 
    \item AI should \textbf{accommodate diverse learning needs}. An algebra teacher for disabled students stressed that (T4) stated “every kid's different. We need to make the learning fit each one.” 
    \item AI should \textbf{provide positive reinforcement} to keep students motivated. T4 stated “[students] get frustrated easily if it’s too hard.” A kindergarten teacher for visually impaired students also stated (T5) “encouragement and positive feedback make a big difference.” A well-designed AI system should offer guidance and support.
\end{itemize}

\begin{figure}[H]
    \centering
    \includegraphics[width=0.4\textwidth]{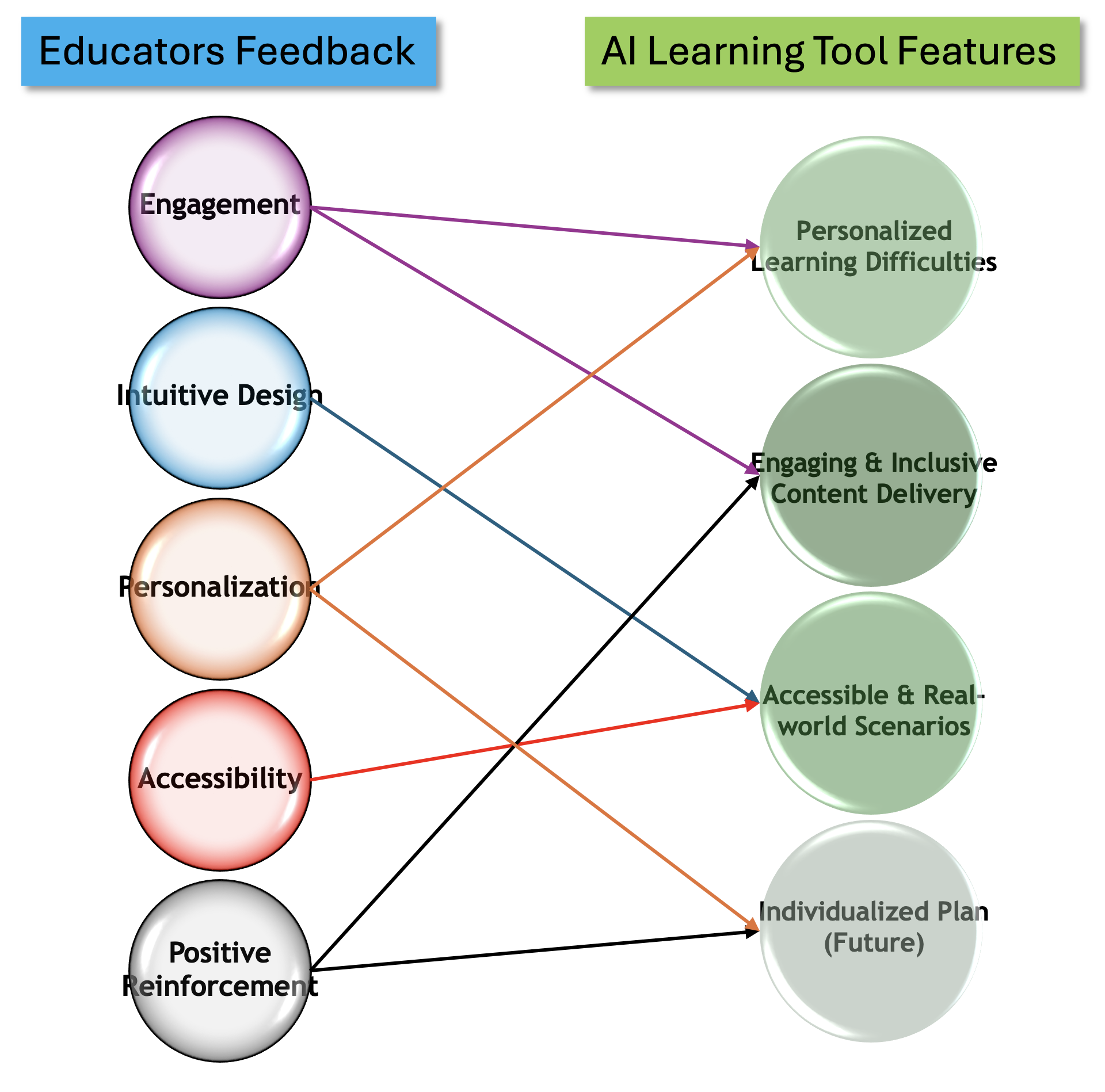}
    \caption{Educators Feedback and Audemy Features Map}
\end{figure}

\subsection{Ethical Considerations in AI for Education}
Audemy prioritizes student data security and privacy through the following measures:

\begin{itemize}
    \item End-to-end encryption protects data during transmission and while stored on secure servers, which prevents unauthorized access.\cite{ibm_encryption}
    \item Voice inputs and learning progress data are separated from personally identifiable information unless explicit consent is given, reducing risks of data breaches and unauthorized profiling.\cite{velotix_anonymization}
    \item Clear guidelines on data usage are in place. Educators and students have transparency on data usage and control over permissions, for example options to request data deletion. 
   
\end{itemize}
\hyphenpenalty = 9000
\section{Future Considerations for Accessible Learning AI}

\subsection{AI Empathy}
One of the current limitations of Audemy is its ability to interpret complex emotional states or make context-sensitive decisions based on audio cues. Future research can focus on incorporating AI-driven emotional assessment techniques. For example, AI could analyze tone, response time, and text input to infer emotional states. If a student sounds frustrated or takes longer to answer, the AI could adjust lesson pacing. In a similar way, if AI detects signs of stress through the audio, such as repeated incorrect responses, it could simplify instructional content. These developments would create a more empathetic, human-like AI tutor.

\subsection{Personalized Educator Resources}

AI-driven student progress tracking could help teachers learn more about student performances through identifying areas where students excel and where they need additional support. Features like automated progress reports and difficulty level tracking could help teachers build individualized learning plans.

\section{Conclusion}
This paper has examined the role of AI in creating personalized, accessible learning experiences for blind and visually impaired (BVI) students through Audemy. The platform uses AI-driven adaptive learning, real-time feedback, and conversational AI to increase engagement and learning outcomes. Accessibility challenges and ethical considerations such as data privacy and educational equity were also addressed. While challenges remain, particularly in emotional responsiveness and tactile feedback, the integration of AI in accessibility education presents an opportunity to make learning for BVI students more personalized and inclusive.

\bibliographystyle{ACM-Reference-Format}
\bibliography{references}

\end{document}